\documentclass{article}
\usepackage{spconf,amsmath,graphicx,mathtools,amsfonts}
\usepackage{MnSymbol} 
\usepackage{subfig}
\usepackage{diagbox}
\usepackage{hyperref}
\usepackage{enumitem}
\setlist{nosep, leftmargin=14pt}

\usepackage{mwe} 

\def\psz{0.19\columnwidth}

\newcommand{\norm}[1]{\left\lVert#1\right\rVert}
\newcommand{\ngrad}[1]{\ensuremath{\vec{n}(#1)}}
\newcommand{\ngradsub}[2]{\ensuremath{\vec{n}_{#2}(#1)}}
\newcommand{\cc}[2]{\ensuremath{(#1 \thinstar #2)}}
\DeclareMathOperator*{\argmax}{arg\,max}

\newcommand{\voxelunit}{\ensuremath{\textit{\,vx}}}

\title{Cross-Sim-NGF: FFT-Based Global Rigid Multimodal Alignment of Image Volumes using Normalized Gradient Fields}
\name{Johan \"{O}fverstedt, Joakim Lindblad, Nata\v{s}a Sladoje}
\address{Department of Information Technology, Uppsala University, Uppsala, Sweden}
\begin{document}
%
\maketitle
\begin{abstract}
Multimodal image alignment involves finding spatial correspondences between volumes varying in appearance and structure. Automated alignment methods are often based on local optimization that can be highly sensitive to their initialization. We propose a global optimization method for rigid multimodal 3D image alignment, based on a novel efficient algorithm for computing similarity of normalized gradient fields (NGF) in the frequency domain. We validate the method experimentally on a dataset comprised of 20 brain volumes acquired in four modalities (T1w, Flair, CT, [18F] FDG PET), synthetically displaced with known transformations. The proposed method exhibits excellent performance on all six possible modality combinations, and outperforms all four reference methods by a large margin. The method is fast; a 3.4Mvoxel global rigid alignment requires approximately 40 seconds of computation, and the proposed algorithm outperforms a direct algorithm for the same task by more than three orders of magnitude. Open-source implementation is provided.
\end{abstract}
\section{Introduction}

Multimodal image alignment (also known as registration) involves finding correspondences between images with varying degree of difference of appearance and structure \cite{zitova2003image}, often with the goal of combining the characteristics of each modality via image fusion. Alignment of large displacements is particularly challenging since correspondences to be inferred are far apart, requiring global contextual and spatial information.

While many methods for monomodal alignment exist, much fewer general-purpose multimodal alignment methods exhibiting  high performance and robustness (w.r.t. the choice of the modalities to be aligned) are available. Common approaches include local optimization methods based on mutual information (MI) \cite{viola1997alignment, kleinElastixToolboxIntensityBased2010}, as well as normalized gradient fields (NGF) \cite{haber2006intensity}, and representation extraction techniques based on local self-similaries \cite{heinrichMINDModalityIndependent2012}, as well as Deep Feature Learning \cite{pielawskiCoMIRContrastiveMultimodal2020,islam2021deep}. Recently, a global alignment method based on the cross-mutual information function (CMIF) was presented \cite{ofverstedt2021fast}.

\begin{figure}[t]
    \centering
    \includegraphics[width=\columnwidth]{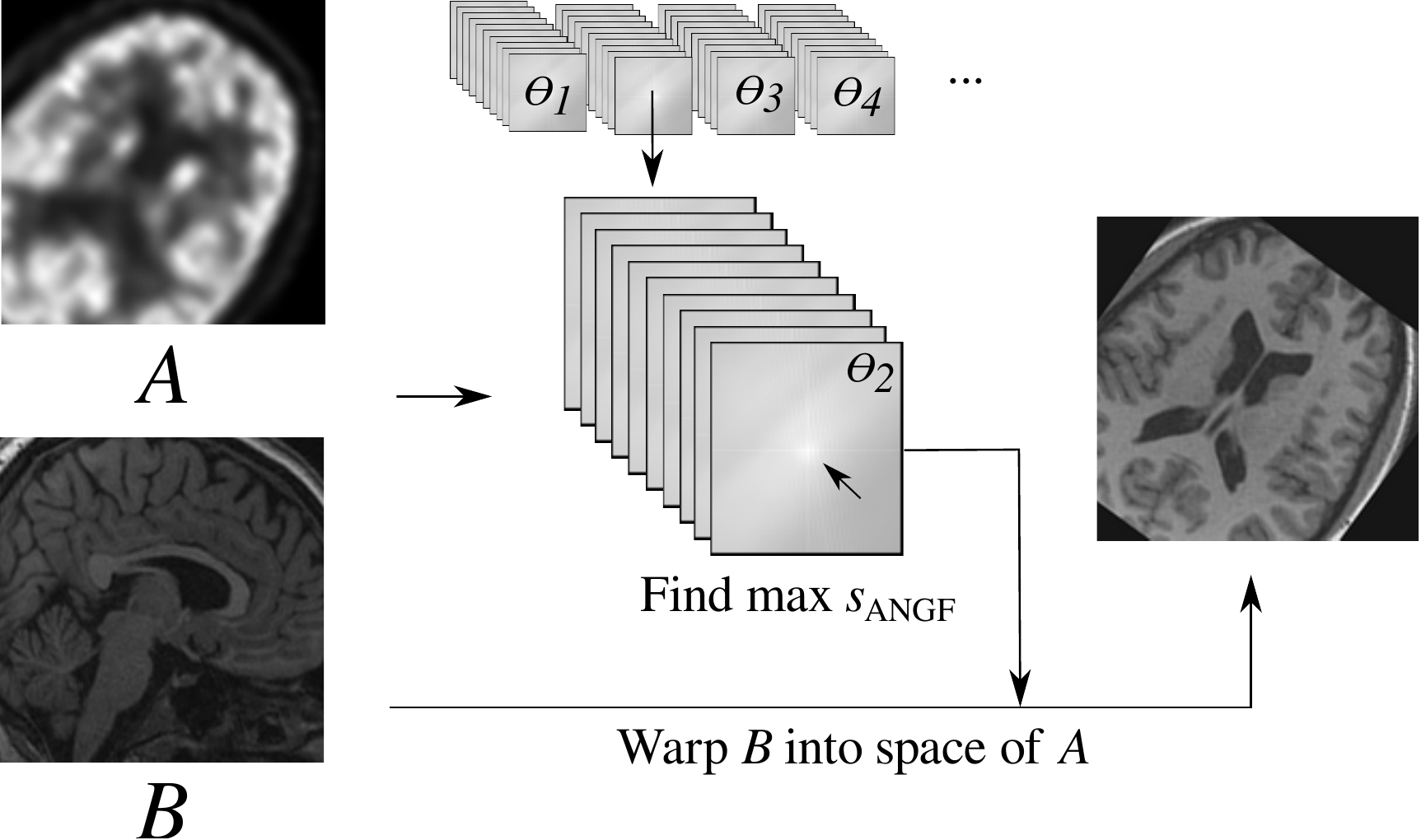}
    \caption{Main idea of the proposed global alignment method. Input: Two image volumes of modalities [18F] FDG PET, and T1 weighted MR, as $A$ and $B$ respectively (displayed as slices). For a number of random 3D rotations $\mathbf{\theta}$, the similarity measure $s_{\text{ANGF}}$ between the masked normalized gradient fields is computed efficiently for all 3D displacements; finally, the sought transformation is found as the maximum of $s_{\text{ANGF}}$.}
    \label{fig:fig1}
\end{figure}

We here propose a new global alignment method based on NGF, that is fast and shows excellent performance on a rigid multimodal 3D medical image alignment task, compared to local optimization-based MI and NGF, and global alignment-based CMIF, as well as a version of the proposed method using a related (but distinct) existing similarity measure, on 6 pairs of modality combinations.  
Figure~\ref{fig:fig1} illustrates the general idea of the method.

A fast and user-friendly PyTorch-based implementation of the method with open-source code is available at \href{https://github.com/MIDA-group/cross_sim_ngf}{http://github.com/MIDA-group/cross\_sim\_ngf}.

\section{Background}
Here we recall the most relevant aspects of NGF and methods for computing them.

The (regularized) normalized gradient field \cite{haber2006intensity}, for image $A$ at point $x$, is given by
\begin{equation}
\ngrad{x; A} = \frac{\nabla{A(x)}}{\sqrt{\norm{\nabla{A(x)}}_2^2 + \epsilon^2}},
\end{equation}
where $\epsilon$ is a small constant introduced to reduce the impact of gradients with very small magnitude, and avoid division by zero. In this work we use $\epsilon=10^{-5}$ (selected empirically). The main assumption of NGF is that parts of images acquired by different modalities are in correspondence when the directions of their intensity changes are parallel or anti-parallel. 
A similarity of NGF (SNGF) based on the squared dot-product of the elements of the NGF can be defined as
\begin{equation}
\begin{split}
s_{\text{NGF}}(x; A, B) = {\langle \ngrad{x; A}, \ngrad{x; B} \rangle}^2.
\end{split}
\label{eq:dngf}
\end{equation}
This measure approaches $1$ if the two vectors are parallel or anti-parallel and $0$ if the two vectors are orthogonal. 

Orientation correlation (OC) and squared orientation correlation (SOC) offer an efficient way of computing SNGF of 2D images for all discrete displacements, where 2D gradients are represented as complex numbers \cite{fitch2002orientation}. An efficient algorithm which utilizes log-polar Fourier Transform is proposed for OC-based alignment w.r.t. rotation and scaling in 2D \cite{tzimiropoulos2010robust}.
NGF also appeared as a component of an objective function combined with mutual information to enhance the performance of multimodal image alignment tasks \cite{pluim2000image}.

NGF cross-correlation was extended to 3D in the context of medical template matching \cite{fotin2012normalized}, by utilizing a modified version of \eqref{eq:dngf} where squaring of the dot-product is omitted:
\begin{equation}
    s_{\text{US-NGF}}(x; A, B) = {\langle \ngrad{x; A}, \ngrad{x; B} \rangle}.
\label{eq:usngf}
\end{equation}
By observing three separable components of the unsquared dot-product, the authors formulated an algorithm for efficiently computing the measure for all discrete displacements using cross-correlation in the frequency domain. Additionally, in \cite{fotin2012normalized} it was observed  that smoothing of the images is important to enhance the efficacy of the method. Eq.~\eqref{eq:usngf}, similarly to (the unsquared) OC \cite{fitch2002orientation}, exhibits many useful properties, such as invariance to contrast and absolute intensity levels, being sensitive only to the orientation of the images. For many multimodal scenarios (unlike the monomodal template matching scenario in \cite{fotin2012normalized}), one part of a specimen appears dark in one modality and bright in another, and it can thus be highly detrimental to the alignment performance if the objective function counts them as misaligned.

\section{Method}

\subsection{Algorithm for Fast Computation of Similarity of NGF in the Frequency Domain}
\label{sec:algorithm1}

In \cite{haber2006intensity}, the point-wise contributions of $s_{\text{NGF}}$  \eqref{eq:dngf} are aggregated by summation. 
Here, we formulate a scaled similarity measure which is applicable to selected sub-regions of the images, while masking out the remaining parts of the finite rectangular domains. 
The similarity of average NGF is
\begin{equation}
\begin{split}
s_{\text{ANGF}}(A, B; M_A, M_B) =\\ 
\frac{1}{\sum_x{M_A(x)M_B(x)}}\sum_{\mathclap{x}}{M_A(x)M_B(x)s_{\text{NGF}}}(x; A, B)\,,
\end{split}
\end{equation}
where masks $M_A$ and $M_B$ are indicator functions on the domain.
Based on computed $s_{\text{ANGF}}$ for displaced images $B$, we define the \emph{cross similarity of NGF}
\begin{equation}
\begin{split}
\label{eq:csngf}
\text{CSNGF}(\chi; A, B, M_A, M_B) =\\
\frac{1}{N(\chi)}\sum\limits_{x}{M_A(x)M_B(x+\chi)s_{\text{NGF}}}(x; A(x), B(x+\chi)),
\end{split}\!
\end{equation}
where $\chi$ is a discrete translation, and where $N(\chi)$ is the number of overlapping voxels (where the masks intersect) as a function of $\chi$ and can be computed as the cross-correlation between the two mask images $N(\chi)=\cc{M_A}{M_B}(\chi)$, analogously to how masks are incorporated in CMIF \cite{ofverstedt2021fast}.
To compute CSNGF in the frequency domain for all $\chi$ in 3D, we rewrite \eqref{eq:dngf}, by expanding the squared dot-product and obtain
\begin{equation}
\begin{split}
s_{\text{NGF}}(x; A, B) &= %
\sum_{\mathclap{i=1}}^{3}\Big(\ngradsub{x; A}{i}^2{\ngradsub{x; B}{i}}^2 +\\ 2&\sum_{\mathclap{j=i+1}}^{3} \ngradsub{x; A}{i}\ngradsub{x; A}{j}\ngradsub{x; B}{i}\ngradsub{x; B}{j} \Big)
\end{split}
\end{equation}
which consists of 6 separable parts, each computable independently using cross-correlation. The cross-correlations
\begin{equation}
\cc{\ngradsub{\cdot\,; A}{i}^2}{\ngradsub{\cdot\,; B}{i}^2}\,,
\end{equation}
\begin{equation}
\cc{(\ngradsub{\cdot\,; A}{i}\ngradsub{\cdot\,; A}{j})}{(\ngradsub{\cdot\,; B}{i}\ngradsub{\cdot\,; B}{j})}
\end{equation}
are efficiently computed in the frequency domain (and combined via summation before application of the inverse transform). Computing CSNGF involves 14 real-valued FFTs (6 per image plus 1 mask for each image) and 2 inverse FFTs, and a few element-wise complex sums and multiplications per voxel. This general approach is applicable to images of arbitrary dimension; for clarity, we here only observe the 3D case.

\subsection{Method for Global 3D Rigid Alignment}

The efficient algorithm for computing the CSNGF for all $\chi$ provides direct means of global optimization of $s_{\text{ANGF}}$ w.r.t. axis-aligned shifts. To enable global optimization w.r.t. rigid transformations, we adopt a hybrid approach where the rotation parameters $\theta$ (represented as Euler angles) are explored with random search and the optimal displacements for a given rotation are then located by computing $\argmax_\chi \text{CSNGF}$.

For each randomly selected rotation vector $\theta$, the corresponding transformation $T_{\theta}$ is applied to the floating image $\hat{B}=B \circ T_{\theta}$ using trilinear interpolation.  $\ngrad{\cdot\,; B \circ T_{\theta}}$ is computed, followed by computation of CSNGF for all $\chi$ satisfying a user-selected amount of minimum overlap $\gamma$, and using a suitable zero padding scheme (following \cite{ofverstedt2021fast}). We use 
$\gamma=0.5$ everywhere in this work.

Let $\bar{\Theta}$ denote a set of starting points of 3D rotations, $u$ denote the maximum allowed step, $\bar{\theta} \sim U(\bar{\Theta})$, where $U(\bar{\Theta})$ denotes the uniform discrete distribution from the set $\bar{\Theta}$, and $r_x \sim \mathbb{U}(-u, u)$, $r_y \sim \mathbb{U}(-u, u)$, $r_z \sim \mathbb{U}(-u, u)$. We then sample a new 3D rotation $\theta = \left\{\bar{\theta}_x + r_x, \bar{\theta}_y + r_y, \bar{\theta}_z + r_z\right\}$, where $\bar{\theta}_x$ denotes a rotation around the $x$-axis, $\bar{\theta}_y$ around the resulting $y$-axis, and $\bar{\theta}_z$ around the resulting $z$-axis.

Both for purposes of speed and widened convergence region around the global maximum, we employ a multi-stage, Gaussian pyramid scheme with smoothing and downsampling of the images, with $m$ levels.
Let $\bar{\Theta}_{k}$, for level $k$, comprise the best rotations w.r.t. the similarity measure from the previous stage (or user-selected starting points for the first stage). In addition to random sampling relative to these rotations, we also compute CSNGF for all elements of $\bar{\Theta}_{k}$, to avoid missing an already located good solution due to randomness.

The pyramid scheme is parameterized by smoothing parameters $\mathbf{\sigma}=(\sigma_1, \dots, \sigma_m)$, downsampling factors $\mathbf{d}=(d_1, \dots d_m)$, maximum allowed steps $\mathbf{u}=(u_1, \dots u_m)$, number of random rotations $\mathbf{a}=(\text{a}_1, \dots \text{a}_m)$, and number of starting points $\mathbf{p}=(p_1, \dots, p_m)$. 
For all experiments using this framework, we use  $\mathbf{d}=(4, 2, 2, 1)$, $\mathbf{s}=(180, 30, 10, 0)$, $\mathbf{a}=(5000, 3000, 300, 0)$, $\mathbf{p}=(1, 20, 3, 1)$, with the initial starting point $(0, 0, 0)$.

\section{Data}
\label{sec:data}

The empirical evaluation of the proposed method is based on the CERMEP-IDB-MRXFDG dataset \cite{merida2021cermep}, available upon request from the authors. The dataset consists of images of brains of 33 subjects acquired by 4 different modalities: T1 weighted MR, Flair MRI, Computed Tomography (CT), [18F] FDG PET, all mapped to the standard MNI space, thus providing ground-truth for image alignment method evaluation, and a possibility to consider 6 different combinations of modalities, enabling evaluation of the methods in terms of generality and robustness.

\subsection{Construction of The Evaluation Dataset}

\begin{figure}
    \centering
    \subfloat[][Ref1: PET]{\includegraphics[width=\psz]{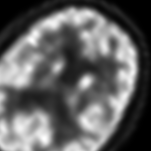}\label{subfig:data00}}\hspace{3mm}
    \subfloat[][Ref2: T1]{\includegraphics[width=\psz]{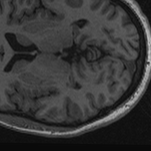}\label{subfig:data10}}\hspace{3mm}
    \subfloat[][Ref3: Flair]{\includegraphics[width=\psz]{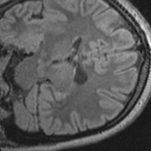}\label{subfig:data20}}\hspace{3mm}
    \subfloat[][Ref4: CT]{\includegraphics[width=\psz]{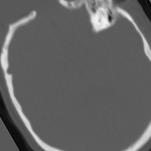}\label{subfig:data30}} \\[1ex]
    \subfloat[][Flo1: T1]{\includegraphics[width=\psz]{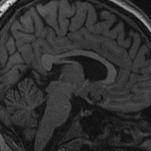}\label{subfig:data01}}\hspace{3mm}
    \subfloat[][Flo2: Flair]{\includegraphics[width=\psz]{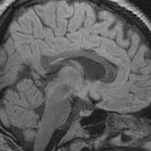}\label{subfig:data11}}\hspace{3mm}
    \subfloat[][Flo3: CT]{\includegraphics[width=\psz]{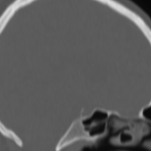}\label{subfig:data21}}\hspace{3mm}
    \subfloat[][Flo4: PET]{\includegraphics[width=\psz]{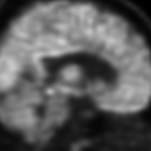}\label{subfig:data31}}\\[1ex]
    \subfloat[][GT1: T1]{\includegraphics[width=\psz]{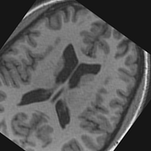}\label{subfig:data02}}\hspace{3mm}
    \subfloat[][GT2: Flair]{\includegraphics[width=\psz]{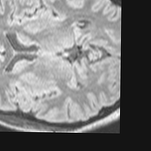}\label{subfig:data12}}\hspace{3mm}
    \subfloat[][GT3: CT]{\includegraphics[width=\psz]{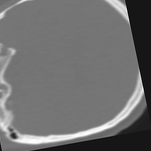}\label{subfig:data22}}\hspace{3mm}
    \subfloat[][GT4: PET]{\includegraphics[width=\psz]{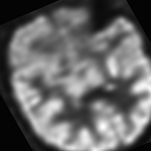}\label{subfig:data32}}
    \caption{Sample slices of 3D image pairs from the evaluation dataset generated from the CERMEP-IDB-MRXFDG dataset \cite{merida2021cermep}. (a-d) the reference (transformed) images and (e-h) the floating images. Image (e) is to be registered to (a); (f) to (b), (g) to (c) and (h) to (d). The bottom row shows the ground-truth (GT) of each floating (Flo) image aligned to the corresponding reference (Ref) image.}
    \label{fig:data}
\end{figure}

For each of the twelve (ordered) pairs of modalities (six unordered modality combinations), and for each of the first 20 subjects (leaving the last 13 for testing), we randomly (uniformly) sample a 3D rotation $\theta$, and an axis-aligned shift $\chi_i \in \left[-30\voxelunit, +30\voxelunit\right]$ for each axis $i$. These transformations are applied using inverse mapping and bicubic interpolation, with zero-padding, to the first image volume of each pair. The transformed image is taken as reference image and the untransformed image as floating image in the alignment task. Finally, a block of size $151 \times 151 \times 151 \voxelunit$ (c.f. original size $207 \times 243 \times 226$) at the center of the volume is extracted, retaining most of the content of interest, while omitting most of the background. Examples are shown in Fig.~\ref{fig:data}.

\section{Performance Analysis}

\subsection{Multimodal Brain Image Volume Alignment}

We first evaluate the performance of the proposed method w.r.t. the accuracy of global multimodal alignment of 3D images, compared to several widely used general-purpose methods and a recent global alignment method based on CMIF.

\subsubsection{Method Selection and Experimental Setup}

Let {USNGF} refer to an alignment method
based on CSNGF, with $s_{\text{NGF}}$ replaced by $s_{\text{US-NGF}}$ in \eqref{eq:csngf}. {USNGF} is included in this study as the closest related method to our proposed method CSNGF, with the aim to evaluate the advantage to the proposed algorithm in relation to what could be achieved with the algorithm in \cite{fotin2012normalized}. The recent CMIF-based global alignment method \cite{ofverstedt2021fast}, which exhibited excellent performance and outperformed several recent Deep Learning methods (including \cite{pielawskiCoMIRContrastiveMultimodal2020}) on multiple biomedical datasets, is another reference method.
All the selected global optimization methods are implemented in Python/PyTorch \cite{paszke2019pytorch} with CUDA/GPU-acceleration. We also compare with local optimization-based methods using MI and NGF as objective functions, relying on open-source implementations Elastix \cite{kleinElastixToolboxIntensityBased2010} and AIRLab \cite{sandkuhlerAIRLAB2018} respectively.

We select the mean Euclidean distance between he corresponding corner points of the extracted block and its version after the performed (recovered) alignment as a displacement measure, denoted $d_E$. We consider an alignment successful if $d_E < 5\voxelunit$, which is $2\%$ of the length of the diagonal of the blocks, rounded to the nearest integer.

We include both {USNGF} and ``USNGF-'', which designates the method {USNGF} with an intensity-inverted floating image, introduced with an aim to investigate the sensitivity of USNGF to the sign of a gradient \cite{fitch2002orientation}.

For CMIF we use $k=16$ (for the $k$-means clustering), and $\mathbf{\sigma}=(3.0, 1.5, 1.0,0.0)$. For NGF, USNGF (and USNGF-), we use $\mathbf{\sigma}=(5.0, 3.0, 2.0, 1.5)$.
For local optimization MI (LO-MI) \cite{viola1997alignment,kleinElastixToolboxIntensityBased2010}, we use 6 pyramid levels, the Adaptive Stochastic Gradient Descent optimizer \cite{kleinAdaptiveStochasticGradient2008}, 4096 maximum iterations for each level. For local optimization NGF (LO-NGF) \cite{haber2006intensity}, we use 5 pyramid levels, the ADAM optimizer, iteration counts according to the schedule (4096, 4096, 1024, 100, 50), with downsampling factors (16, 8, 4, 2, 1) and Guassian smoothing parameters (15.0, 9.0, 5.0, 3.0, 1.0), with learning-rate 0.01. Trilinear interpolation is used.

\subsubsection{Results}

The results of the evaluation of the 6 considered methods on the multimodal brain image dataset are presented in Tab.~\ref{tab:resultstable}. The proposed method provides overall excellent performance, and is the best choice for all observed modality combinations. Most of the competitors show generally poor performance, completely failing on one or more modality combinations.  

\begin{table}[]
    \centering
    \caption{Image alignment performance presented in terms of success-rate, where the threshold of success is set to $5\voxelunit$. The modality names are abbreviated in the headings (T: T1, F: Flair, C: CT, P: [18F] FDG PET). The results for the two directions for each modality combination is aggregated. }
    \label{tab:resultstable}
    \resizebox{0.85\columnwidth}{!}{
    \begin{tabular}{l|cccccc}
        \diagbox{Method}{Modalities} & T/F & T/C & T/P & F/C & F/P & C/P\\ \hline
        LO-MI & 0.05 & 0.025 & 0.075 & 0.025 & 0.1 & 0.075 \\
        LO-NGF & 0.025 & 0.00 & 0.00 & 0.00 & 0.00 & 0.00\\\hline
        CMIF & 0.675 & 0.30 & 0.325 & 0.80 & 0.85 & 0.525 \\
        USNGF & 0.225 & 0.00 & 0.00 & 0.00 & \textbf{0.925} & 0.10 \\
        USNGF- & 0.00 & 0.275 & 0.00 & 0.00 & 0.00 & 0.00 \\
        \textbf{CSNGF} & \textbf{1.00} & \textbf{0.95} & \textbf{0.925} & \textbf{0.90} & \textbf{0.925} & \textbf{0.95} \\
    \end{tabular}
    }
\end{table}

\begin{figure}
    \centering
    \includegraphics[width=0.95\columnwidth]{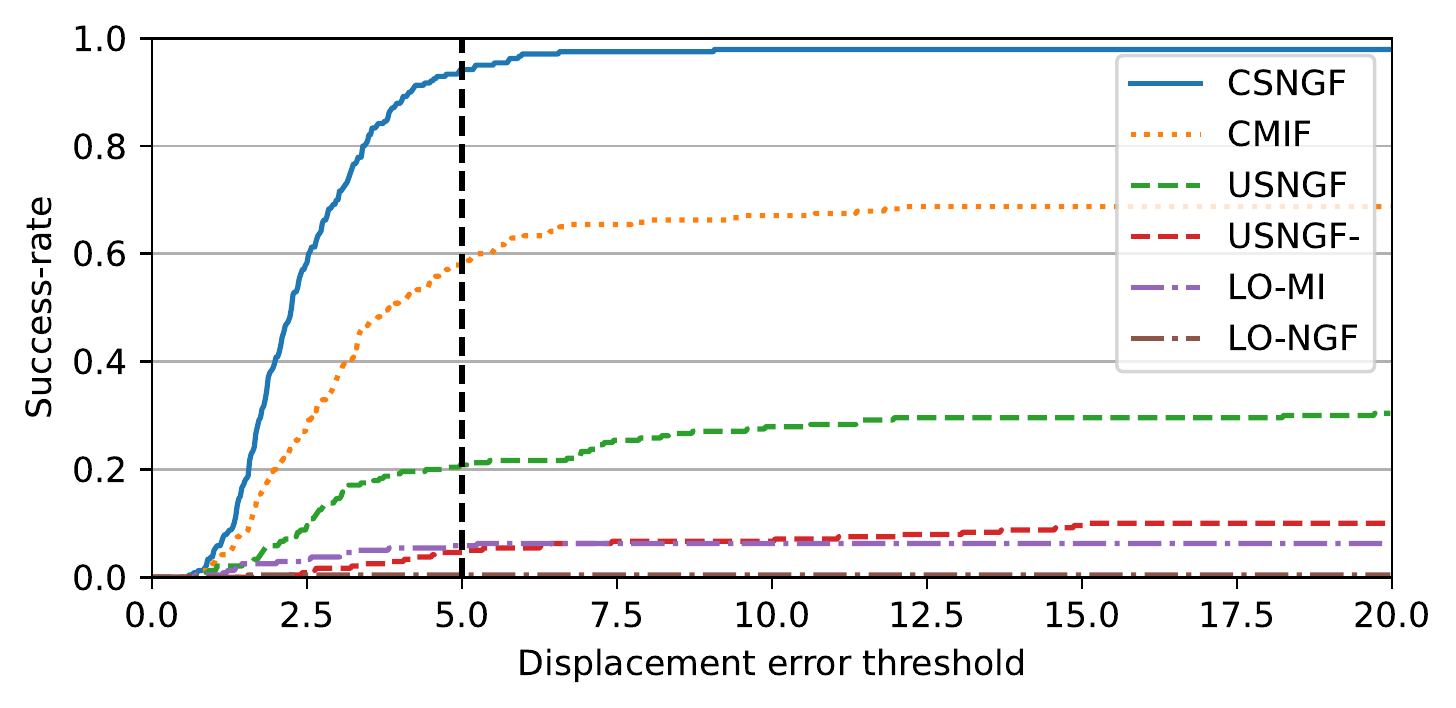}
    \caption{The success-rate of each considered method as a function of the acceptable displacement error $t$ (fraction of the 240 alignments where $d_E<t$); the results for all modality combinations are aggregated. Up and to the left is better.}
    \label{fig:cumulative}
\end{figure}

For global alignment processes, near-successes are also of interest, since those solutions may be refined with a local optimization method; therefore we plot the full cumulative distribution up to the threshold $d_E<20$ as Fig.~\ref{fig:cumulative}.

\subsection{Time Analysis}

The evaluation of time complexity has two parts; (i) comparison of run-times of the observed rigid registration methods, and (ii) comparison of run-times of the proposed Cross-Sim-NGF algorithm with a direct (not FFT-based) approach.

The reported results for the global methods (CMIF, USNGF, CSNGF) are obtained on a Nvidia GeForce RTX 3090. The selected implementations of the local optimization methods are run on the CPU. They are not GPU-accelerated, and have far less inherent parallelism (due to dependency between each iteration), which makes comparisons difficult; their run-times are still included here for reference.

Both the FFT-based algorithm and the direct method are implemented in Python, and use PyTorch to utilize GPU-acceleration; the direct method is implemented as a nested loop over all valid axis-aligned shifts $\chi$, and direct computation of the squared dot-products.

The run-times of all methods for a full alignment process are presented in Tab.~\ref{tab:runtimes}. Comparison of the run-times of the FFT-based algorithm and the direct method, as a function of image size, is presented in Tab.~\ref{tab:runtimes2}. We observe that for size $128$, the here proposed algorithm is approximately $6275$ times (more than three orders of magnitude) faster.

\begin{table}
\centering
\caption{Run-times of the considered methods in seconds.}
\label{tab:runtimes}
\resizebox{0.85\columnwidth}{!}{
\begin{tabular}{l|cc|ccc}
Method & LO-MI & LO-NGF & CMIF & USNGF & \textbf{CSNGF} \\\hline
Run-time &  $52$ & $61$ & $569$ & $33$ & $41$
\end{tabular}
}
\end{table}

\begin{table}
\centering
\caption{Run-time comparison of FFT-based Cross-Sim-NGF and a direct algorithm for computing CSNGF on cube image volumes as a function of size (expressed as side-length).}
\label{tab:runtimes2}
\resizebox{0.85\columnwidth}{!}{
\begin{tabular}{l|ccccc}
\diagbox{Method}{Size} & 8 & 16 & 32 & 64 & 128 \\\hline
Direct algorithm & 0.129 & 0.557 & 3.537 & 27.07 & 502.4\\
\textbf{FFT-based alg.} & $\mathbf{0.002}$ & $\mathbf{0.002}$ & $\mathbf{0.002}$ & $\mathbf{0.008}$ & $\mathbf{0.088}$
\end{tabular}
}
\end{table}

\section{Conclusion}

We propose a novel approach to use NGF for global rigid multimodal 3D medical image alignment. We confirm both its great performance and its high efficiency, benefiting from FFT processing and the parallel compute capability of GPUs. The method does not require or use any training (data), a significant advantage for (bio)medical applications~\cite{islam2021deep}.

\section{Compliance with ethical standards}
\label{sec:ethics}

This research study was conducted retrospectively using human subject data made available by \cite{merida2021cermep}. Ethical approval was not required as confirmed by the license attached with the data.

\section{Acknowledgments}
\label{sec:acknowledgments}

We wish to thank Ines M\'{e}rida and team for providing the CERMEP-IDB-MRXFDG dataset.
The authors declare that there are no possible conflicts of interest related to this research and the contents of this manuscript.

\bibliographystyle{IEEEbib}
\bibliography{strings,refs}

\end{document}